# White Paper

*Human-AI Collaboration in Conflict Analysis: Text Classifier Development with Peacebuilders*

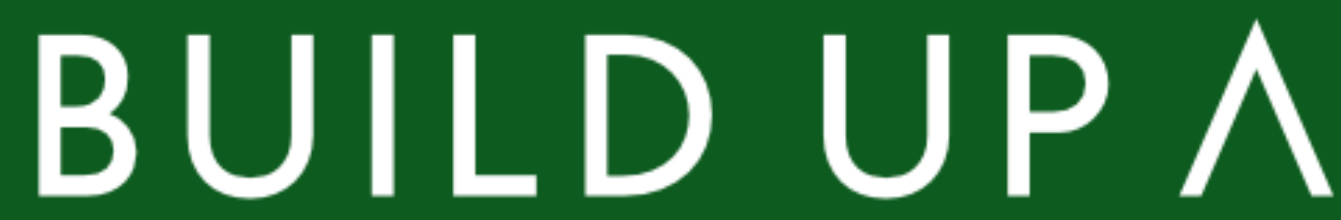

# About this report

This paper documents a collaborative research process involving peacebuilders and data scientists in Kenya and Sudan to develop AI-based text classifiers for monitoring online polarization and hatespeech. The method describes a participatory annotation process in which practitioners and domain experts contributed to problem definition, annotation design, iterative validation, and model evaluation. Fine-tuned BERT-based classifiers were trained on collaboratively annotated datasets and evaluated against held-out test sets. In each case, the models produced enhanced contextual alignment, reduced misclassification driven by cultural nuance, and increased practitioner ownership of AI tools. The resulting models (Kenya-polarization and Sudan-hate speech) are open-source and accessible via HuggingFace. The study contributes empirical evidence that participatory AI development can simultaneously improve technical robustness, contextual validity, and normative alignment in sensitive humanitarian domains.

## Authors & Acknowledgments:

Build Up and DataValuePeople co-lead the Automatic Classifiers for Peace (ACfP) initiative, a joint effort to develop open-source AI tools for conflict analysis in partnership with peacebuilding practitioners. This research was conducted under that initiative across two contexts, Kenya and Sudan, with practitioner cohorts recruited and engaged as co-researchers throughout the classifier development process. The work was supported by the Notre Dame IBM Technology Ethics Lab, with additional institutional support from the PeaceTech and Polarization Lab at the University of Notre Dame. This white paper is accompanied by a fuller academic treatment of the methodology and findings, forthcoming for journal submission.

## To cite:

*For questions about this white paper or to request support, please contact [team@howtobuildup.org](mailto:team@howtobuildup.org)*

# INTRODUCTION

There are three interconnected issues at the intersection of AI use and development and peacebuilding practice that motivate this research. We can distinguish these as problems of fit, of co-design, and of participation.

First, there is a misfit between existing tools and peacebuilding goals. The growth of computational approaches to digital media analysis has produced a landscape of tools designed primarily for marketing, brand monitoring, and audience segmentation. When peacebuilding organizations with sufficient resources have turned to these commercial platforms, they have found that their affordances fall short of what peacebuilding work requires. Affective polarization, hatespeech, intimidation, or incitements to violence manifest differently and dynamically across contexts and languages than existing toxicity taxonomies can encode. The result is a set of tools that cannot process the type of social media data a conflict practitioner needs to see, especially in low-resource language contexts.

A second related issue is the absence of tailor-made applications co-designed with practitioners. Models developed for generic use, with generic or synthetic training data, cannot capture the contextual expertise and grounded objectives that co-design with practitioners can. Affective polarization, hatespeech, intimidation, and incitements to violence manifest in more contexts and languages than existing toxicity taxonomies and capture. In conflict settings, coded language also shifts deliberately to evade detection. An absence of co-design also undermines utilization. When practitioners have no roles in development, they are provided with tools that they may not understand or trust enough to integrate into workflows.

Finally, there is tension between AI's characteristic promises of efficiency, scale, and synthetic aggregation and normative commitments of peace researchers and practitioners. Engagement and partnership with affected communities is constitutive of the work itself, and there is a risk when communities are represented through model outputs rather than engaged in deliberative processes. This paper takes seriously the concern that poorly designed AI integration can hollow out those participatory commitments that give peacebuilding its legitimacy. The question is to what extent and to what effect model development can be a participatory process itself.

The study's participatory approach to developing custom text classifiers extends efforts to create hatespeech detectors through NGO-AI collaborations (Tekiroglu et al. 2020; Chung et al. 2021) and responds to calls for critical engagement with digital technologies in peace processes (Hirblinger et al. 2023). Previous research has also shown that generative AI can successfully support classification models (Törnberg 2023; Møller et al. 2023; Gilardi et al. 2023; Rathje 2023). Still, it is unknown to what extent and with what level of training or prompt engineering this could include conflict-related data. By focusing on reducing annotation time and increasing accessibility, this research addresses the need for efficient, context-sensitive tools in peacebuilding (British Council 2022; Puig Larrauri 2023), thereby helping bridge the gap between AI capabilities and the practical needs of peacebuilding.

There are growing numbers of researchers and practitioners in the field of peacebuilding who are either interested in the uses of AI for improved conflict analysis and response or are compelled by the discourse of inevitability that surrounds it. Mediators and peacemakers working towards the aims of justice, inclusive societies, and accountable institutions recognize the need for digital conflict analysis to identify evidence of divisive behavior online that affects the prospects for peace. The aim of the peacebuilders, researchers, and data scientists co-authoring this paper is to develop tools that help practitioners monitor divisive online discourse at a scale that manual review cannot achieve. The ultimate goal is downstream utility: tools that practitioners can deploy for conflict analysis, interventions, and policy inputs.

## Research Questions

This research documents how early-stage collaboration shapes both human and technical outcomes through two complementary questions:

1. To what extent does active participation in AI-driven text classification affect peacebuilders' awareness of AI capabilities and limitations, attitudes toward AI integration, and willingness to adopt AI technologies in their professional practice?

2. In what ways does peacebuilders' involvement in AI-driven text classification improve the quality and relevance of AI-generated classifications for peacebuilding applications?

By developing and documenting effective practices for integrating practitioner expertise in the use of AI, in this case for text classification, this research provides insights into how AI can augment human capabilities and be responsive to domain-specific needs. The focus on peacebuilding, with its inherent complexities around culture, conflict, and human behavior, offers an ideal case study for developing ethical frameworks that can guide AI deployment in other sensitive contexts. We argue that collaboration that is central at the experimental, iterative classification phase will lead to better deployment outcomes in peacebuilding and comparable fields where practical application is essential.

# METHODOLOGY

### 1.1 Research Design

This study employs an action research methodology that deliberately shifts from the conventional 'data laborer' model of AI development to a co-researcher model, grounded in participatory design and research processes. In the 'data laborer' model, the primary annotators are often not domain experts or practitioners; therefore, their role is confined to labelling data using predefined categories designed by technical or developer teams. The annotators are not involved in the problem formulation process, definition of research goals.

In peacebuilding contexts, this model introduces risks. Conflict-related discussions are highly context and culture-dependent and often employ coded language. Having generic categories and definitions may overlook subtle markers of violence escalation, sarcasm, irony, or historical contexts. This leads to classification systems that risk misalignment with local lived realities that they aim to analyse. In addition, this model is often extractive, meaning that annotators focus more on completion than on contextualization and the accuracy of their inputs. This model also has low uptake because practitioners who use it often don't understand the development process and may perceive it as externally imposed, thereby reducing their willingness to adopt AI technologies, particularly in sensitive fields such as peacebuilding and conflict analysis.

The co-research model, embedded in participatory design, positions practitioners, in this case peacebuilders, as active collaborators. They are involved in problem definition, classification design, and model testing, therefore bringing contextual relevance to the model development process. In this study, peacebuilding practitioners collaborated with the Automatic Classifiers for Peace (ACfP) developers to identify conflict-relevant patterns in social media data, design classification frameworks, evaluate results, and refine approaches, leveraging their expertise and practical needs.

## 1.2 Annotation Process and Participants

The study engaged peacebuilding practitioners in two distinct conflict contexts, Kenya and Sudan. In each case, participant selection was based on their professional experience in monitoring digital harms, conducting conflict analyses, or engaging in content moderation practice. While the overarching methodological framework remained consistent across both contexts, the classification focus varied according to the problem identified by the peacebuilders during the design phase. In Kenya, the cohort included practitioners from civil society organizations, government institutions such as the National Cohesion and Integration Commission (NCIC) and the National Steering Committee (NSC) on Peacebuilding and Conflict Management at the Ministry of Interior, media organisations and content moderation coalitions. In Sudan, practitioners were drawn from independent research institutions and political activist networks, including an open-source investigator specializing in human rights documentation, consultants in election observation and governance, and peace activists from grassroots youth movements. Diversity ensured that the model was not confined to a single institutional domain. In order to assess practitioner outcomes, peacebuilders were interviewed to assess awareness of the model development process, attitudes about the use of AI for peacebuilding, and adoption criteria at the beginning and the close of the development process.

The annotation and model development process followed a structured, multi-step approach designed to reduce bias and improve model accuracy. The process comprised three interlinked steps: (1) development, training, and testing of the annotation process, (2) creation of a randomly selected gold standard holdout dataset, and (3) experimental classifier development and evaluation.

A total of four cycles were implemented, each beginning with collective problem definition, in which practitioners identified specific types of conflict-related content to track and understand. The group then worked together to develop scraping and classification guidelines, annotate the initial datasets, and evaluate how well different AI approaches capture the nuanced ways in which conflict manifests in online spaces. The process included regular reflection sessions in which both technical and domain experts shared insights on what worked, what did not, and how to improve the developer-mediated human-AI collaboration.

The annotation framework began with a clear definition of the classifier based on the target context. In Kenya, the focus was on attitude (affective) polarization, broken down to language that is either stereotyping, dehumanization, deindividuation, vilification, or calls to violence. In Sudan, the focus was hatespeech, adapted to context-specific conflict dynamics. These were translated into a labelling schema, and annotators were trained to ensure everyone had the same understanding of the definitions, while retaining awareness of contextual nuances.

| Variable | Description |
| --- | --- |
| Target classification (e.g., attitude polarisation) | 0 = Not; 1 = Potentially; 2 = Definitely |
| Feature flags (e.g., vilification, dehumanisation) | Presence or absence of specific linguistic features |
| Mark for review | Indicates definitional ambiguity or outlier |

*Table 01: Labelling structure*

Using an online platform, at least three annotators independently labelled each example. This approach increased the total annotation throughput while maintaining reliability safeguards. Posts identified as containing either attitude polarization or hatespeech were subsequently shown to the rest of the annotators for review and confirmation. This ensured positive cases, often rarer, received broader scrutiny. The examples with an equal number of annotations were deemed the "lower" class in the tie; this is a conservative rule that enables results to be reported as a lower bound on attitude polarization or hatespeech.

The process was designed to be iterative and to improve, using an inter-rater reliability (IRR) score computed for each item. After each round, disagreement metrics were computed for each annotated post. Inter-rater reliability statistics were calculated using Krippendorff's α (ordinal distance weighting) for the classifications and Gwet's AC1 for each flag or binary label. A mean value was obtained for each post, indicating whether annotator disagreement was high or low. Examples with high disagreement and items marked for review were discussed in online group sessions, and the harmonised annotations were updated. Selected items were added to subsequent rounds to be re-annotated. This ensured that disagreement was not treated purely as an error, but rather as an indication of how to strengthen the annotation's robustness.

In Kenya, a random sample of 10,633 examples was sampled from a large pool of social media posts collected as part of a larger polarization footprint project. In Sudan, a random sample of 8,746 examples was selected from over 1 million posts collected over a year related to the Sudan conflict. These examples were selected for multi-annotator labelling, starting with a smaller number of posts and gradually increasing the number of examples shown to annotators in subsequent rounds, enabling reliability to be monitored and the annotators' efficiency to improve over time.

## 1.3 Model Development

A subset of the annotated corpus was reserved as a gold standard holdout test set. This dataset provided a stable benchmark for model evaluation and helped mitigate the risks of overfitting and p-hacking by separating the development and evaluation phases.

Classifier development focused on fine-tuning pre-trained transformer architectures, e.g., BERT-type models, for supervised text classification across different hyperparameter settings. The annotated dataset was split into training and test sets. The training set was used for fine-tuning the model, and the test set was used to compute the primary evaluation metric, the F1 Score for a binary distinction between 'definitely' or 'potentially' vs 'not' polarizing (or hatespeech). For the Sudan model, training logs were monitored across epochs to identify points of overfitting as validation loss increased. Training was halted after epoch 5 (table 02).

| Epoch | Training Loss | Validation Loss | Accuracy | Precision | Recall | F1 |
|---|---|---|---|---|---|---|
| 1 | 0.7539 | 0.9286 | 0.9247 | 0.4624 | 0.5000 | 0.4804 |
| 2 | 0.7688 | 0.6723 | 0.9247 | 0.4624 | 0.5000 | 0.4804 |
| 3 | 0.5976 | 0.9373 | 0.9281 | 0.8220 | 0.5361 | 0.5488 |
| 4 | 0.4941 | 0.9430 | 0.9292 | 0.7643 | 0.5915 | 0.6274 |
| 5 | 0.3664 | 0.9327 | 0.9202 | 0.7010 | 0.6415 | 0.6645 |
| 6* | 0.3199 | 1.0339 | 0.9213 | 0.6987 | 0.6079 | 0.6360 |

*Table 02: Training and test dataset split for each model. (* Epoch 6 shown for reference; model training halted after Epoch 5 based on validation loss trajectory.)*

The best-performing fine-tuned transformer model was selected and uploaded to HuggingFace for testing and quality assessment by peacebuilders in both contexts. Tests were also conducted against existing similar models.

# RESULTS

## 2.1 Annotator Outcomes

Technical performance metrics alone are insufficient for evaluating AI tools deployed in high-stakes environments; social legitimacy and user experience are equally consequential. Assessing annotator outcomes is also central to this study's first research question, which asks whether active participation in AI-driven text classification affects peacebuilders' awareness of AI capabilities and limitations, their attitudes toward its integration, and their willingness to adopt it in professional practice. We therefore include structured assessments of annotator experience alongside model performance measures.

Interviews were conducted with all six collaborators in the Kenya cohort and four of the five in the Sudan cohort. Participants ranged in age from 24 to 42 and represented diverse professional backgrounds, including government peace and integration, non-profit program management, journalism and community radio, legal and research analysis, open-source research, software engineering, and election observation consulting. All collaborators engage, to varying degrees, in professional or personal work within peace and justice practice that functions as a shared orientation to the tasks.

Before the annotation process began, the two cohorts entered with meaningfully different baselines. Kenya collaborators were uniformly enthusiastic: all six selected the highest likelihood indicator when asked about future AI use. The dominant sentiment echoed public discourse on inevitability that the technology is here to stay, and peacebuilders need to get on board or get left behind. All had prior consumer-level experience with generative AI tools. Sudan collaborators were more cautious from the outset. Several expressed concern about top-down approaches that serve narrow interests, and at least one argued that AI lacks the empathy required for peace processes and should not substitute for human-centered dialogue in response. Still, within each cohort and often within each collaborator, there were varying attitudes towards AI. Four orientations were present before annotation began:

1. Skepticism, but centered on fit rather than the technology per se. Across both cohorts, the primary concern was whether global AI models could handle locally coded language. Participants noted that hatespeech and polarizing content in their contexts routinely evade generic classifiers because they rely on dialect, historical references, and terminology that shift deliberately to bypass moderation. A related concern was transparency: hidden layers in model logic could produce results that appeared authoritative but were not trustworthy. In Sudan, skepticism also carried a political dimension, with participants questioning who benefits from AI-assisted analysis and under what conditions.
2. Operational interest was strong in both cohorts and centered on scale. Participants described a core problem: the volume of social media data relevant to conflict monitoring is unmanageable for small teams working manually. AI was understood as a way to cover more ground, detect patterns earlier, and be more efficient. "An extra pair of eyes across the country" captured the general aspiration.

3. Professional development motivated participants across both cohorts, though it manifested differently. In Kenya, the emphasis was on building skills and finding a foothold in a rapidly changing technical landscape. In Sudan, it was framed more urgently as not falling behind in a field where capacity gaps already put practitioners at a disadvantage. In both cases, participation in this project was understood explicitly as a learning opportunity that collaborators had opted into, in part, for this motivation.
4. Enthusiasm was most visible in the Kenya cohort, where several participants described AI in expansive terms: as something that "broadens the mind," that could "change the world," or that represented the best professional development opportunity they had encountered. This register was largely absent in Sudan pre-interviews, where optimism, when present, was more qualified and conditional.

In the post-process interviews, willingness to continue using text classification models remained high across both cohorts, with some notable shifts in degree and reasoning. One Sudan collaborator lowered her likelihood score from five to three, citing the poor quality of training data available for conflict-sensitive Arabic contexts and a deepened conviction that models in this space cannot be trusted without rigorous human oversight. Overall, the question had shifted from whether to use AI to how to use it well within existing workflows, and orientations had merged, with most collaborators becoming what might be called skeptical optimists. There were three substantive shifts worth noting:

Pre-interview responses described what AI might do. Post-interview responses drew on specific annotation rounds, particular classification challenges, and direct experience of where the models succeeded and where they did not. Concerns that had been abstract became concrete. The strongest example of this shift was localization. Where contextual fit had been the leading concern at the outset, participants in both cohorts described the process of building something locally grounded as a form of success. "We localized it" captured a sense of ownership and possibility.

The emotional weight of the work came to the fore after the process. Several participants described the cumulative effect of reading large volumes of polarizing or hateful content as heavier than anticipated. This was especially pronounced in the Sudan cohort, where annotators were processing material produced during an active war in their own country, at times in the immediate aftermath of specific atrocities. Participants reflected on how trauma and personal ideology inevitably shape judgment, and how "non-bias" is not an achievable standard for human annotators working on content this close to their own lives.

Finally, skepticism in the post-interviews was more technically specific than before. Concerns shifted from general unease about AI to pointed questions about model reliability in low-resource language contexts, such as the pace at which language shifts relative to model retraining. There were also concerns about dual use, where participants who had built these tools now understood concretely what they could do, and several raised the risk that the same capabilities could be repurposed by governments or malicious actors for surveillance.

Participation did not produce the anticipated increase in technical knowledge as an outcome of the process. Collaborators gained a more nuanced understanding of how AI classification tools can be applied in peacebuilding contexts, but this fell short of functional technical skill-building. While collaborators were substantively engaged in problem definition, taxonomy design, and the iterative refinement of annotation guidelines, they were not regularly engaging with the data scientists running the models or interpreting training outputs. Several collaborators reported understanding the front end of the process well but remained uncertain about how the back end worked. The gap between annotation and model development was experienced as a missed opportunity. Across both cohorts, the majority of collaborators expressed interest in more direct technical engagement, including understanding how labeled data is used in training, how model decisions are made, and how performance is evaluated.

Instead, the most valuable takeaways across the post-interviews were about the process itself. Three levels of reflection emerged consistently: the individual annotator, the annotation group, and the national context in which each annotator worked.

At the individual level, participants reflected on what the process had revealed about their own judgment. Annotating conflict-related content at volume had effects they had not anticipated. Several described the cumulative experience as heavy, and more than one reported that sustained exposure to hateful or polarizing posts affected how they engaged with online spaces afterward. The more substantive realization was about each person's positionality and what that meant for the pursuit of neutrality, with conclusions that prior experience, ideology, and, in the Sudan cohort, proximity to an ongoing war, inevitably shape annotation decisions. Participants did not frame this as a problem to be corrected. Several treated it as a finding about what human annotation can and cannot produce, and by extension, what any classifier trained on it should be expected to do. Finally, some collaborators experienced what they called "reclaiming agency" in digital spaces. By embedding local knowledge into the AI Classifier, the practitioners reclaimed interpretive authority over how their communities were represented in datasets. This was an opportunity to shape how divisive discourse is recognized by countering analytics that flatten local meaning.

At the group level, deliberation sessions convened to resolve labeling disagreements were consistently identified as among the most valuable parts of the project. When annotators with different professional backgrounds reached opposite conclusions on the same post, resolving the disagreement required making the reasoning behind each judgment explicit. Participants reported that this process exposed how institutional position and lived experience shape interpretation in ways that individual annotation does not surface. Working toward a shared label across those differences was described as both more demanding and more informative than annotating independently. Annotation went beyond being a technical task and became a peacebuilding practice in its own right. By engaging peacebuilding practitioners, the process fostered dialogue, contextual reflection, and collective understanding.

At the national level, sustained engagement with large volumes of open social media data exposed participants to content beyond their usual information environments. Several noted that the project produced a more accurate picture of the scale and nature of divisive discourse in their countries than they had held before. It also made concrete a previously understood problem: that language in conflict contexts shifts rapidly and deliberately to evade detection, and that the gap between what a static model can recognize and what a trained local reader can identify is not easily closed. In one example from Sudan, the Arabic term meaning "the wonderful ones" had, by the time of the project, become a vehicle for ethnic hatespeech. This would appear neutral to a generic classifier or even to a native speaker who is not following the online conversation and the terminology shift closely enough.

Owing to these individual, group, and national reorientations, a final finding cuts against assumptions in the literature about the droll experience of data labeling. Annotation at scale is typically experienced as tedious, requiring attention checks and engagement strategies to maintain consistency across large volumes of binary decisions. That was not the dominant experience here. Several said they looked forward to the next batch. One said plainly, "I love it. I love it." Another described being "pretty hooked." Collaborators in this project found the process to be "fun," "very exciting," and "interesting" because it felt personally meaningful. Respondents consistently noted that the work was an "eye-opening" experience that moved beyond technical tasks to trigger deep personal reflections on the "state of society". This journey evoked a wide spectrum of emotions, including surprise at the "reality" of how people speak online, prompting deep personal and collective reflection. This deep engagement forced participants to confront "hard questions" about injustices and engage directly with the volatile divisions and "realities" of their own societies. Ultimately, rather than a tedious run-through of data points, the process was viewed as an "exercise of dialogue and empathy," with annotators describing the labeling process as carrying a "moral responsibility" to get it right. Annotators with a direct stake in the classification problem produced the sustained attention and willingness to deliberate over contested cases in a way that would be difficult to manufacture.

## 2.2 Technical Outcomes

There was a significant class imbalance in the Kenya (attitude polarization) annotated dataset, with only 1.2% (70) of the training data positive, compared to the Sudan (hatespeech) annotated dataset, which had 8.5% (349) of the training data positive. Therefore, the Sudan dataset enabled more stable supervised learning than the Kenya dataset.

| **Context** | **n_total_train** | **n_positive_train** | **n_total_test** | **n_positive_test** |
|---|---|---|---|---|
| Kenya | 5,653 | 70 | 4,980 | 64 |
| Sudan | 4,101 | 349 | 4,645 | 355 |

*Table 03: Training and test dataset split for each model*

The Sudan model achieved a higher overall F1 score (0.673) than the Kenyan model (0.606), with higher precision and recall. In the Kenya model, performance was more constrained, likely because of class imbalance, as it was harder to identify affective polarization compared to hatespeech.

Multiple BERT-based configurations were also trained and compared in the Sudan model. Performance across the runs was relatively similar, but one configuration achieved the highest F1 score and also had the strongest precision and recall. This model was further compared with other open-source, fine-tuned Arabic hatespeech models to determine whether including practitioners in the peacebuilding field in the annotation process improved performance. Although overall accuracy is within a similar range of 0.88-0.92, hate precision and hate recall differ markedly. The Sudan-specific model we trained substantially outperforms generic Arabic hatespeech models and a keyword-based text classifier that the same peacebuilders had been using before.

| Model | Accuracy | Precision (Hate) | Recall (Hate) | Precision (Macro) | Recall (Macro) | F1 (Macro) |
|---|---|---|---|---|---|---|
| rana811/Arabic_HateSpeech_Model | 0.904 | 0.289 | 0.263 | 0.616 | 0.598 | 0.606 |
| Hate-speech-CNERG/dehatebert-mono-arabic | 0.884 | 0.225 | 0.241 | 0.584 | 0.595 | 0.589 |
| Keyword classifier | 0.889 | 0.131 | 0.099 | 0.530 | 0.520 | 0.527 |
| datavaluepeople/Hate-Speech-Sudan-v2 (our model) | 0.920 | 0.425 | 0.388 | 0.688 | 0.660 | 0.673 |

*Table 04. Comparative Model Performance on Sudan Test Set*

For Kenya, several versions of BERT-based models were trained with varying hyperparameters. Some runs achieved similar results while others achieved slightly lower F1 scores; therefore, the final selected Kenya model was the one that showed the most consistent performance on precision and recall, and therefore presented the most reliable overall configuration for detecting affective polarization in the Kenyan dataset.

| Model | Accuracy | Precision (polarization) | Recall (polarization) | Precision (macro) | Recall (macro) | F1 (macro) |
|---|---|---|---|---|---|---|
| Sami92/XLM-R-Large-Polarization-Classifier | 0.978588 | 0.244186 | 0.328125 | 0.617757 | 0.657536 | 0.634566 |
| Catalyst-101/Polarization-Classification-English | 0.888580 | 0.064685 | 0.578125 | 0.529324 | 0.735348 | 0.528447 |
| datavaluepeople/Polarization-Kenya | 0.985924 | 0.370370 | 0.156250 | 0.679803 | 0.576418 | 0.606339 |

*Table 05. Comparative Model Performance on Kenya Test Set*

# DISCUSSION

The findings suggest that participatory involvement of peacebuilders contributes meaningfully to model quality. In the Sudan case, the participatory model (Hate-Speech-Sudan-V2) achieved an F1 score of 0.673, outperforming generic Arabic hatespeech models and keyword-based models. Although the numerical differences are not significant, they reflect improvements in precision and recall. Practitioner input helped clarify culturally specific language, contextualize, and shift forms of expression that generic models failed to capture. In both Sudan and Kenya, domain experts refined what the model was trained to detect, reducing misclassification.

Engaging practitioners from problem definition and annotation increased ownership of the tool. Rather than treating the classifier as an opaque external system, participants developed an understanding of its strengths and limitations. Annotators were not analysing distant or abstract data; they were reading posts embedded in their own lived realities, including discourse surrounding violent events. Despite the psychological strain, participants described the process as a responsibility and, in some cases, as a form of peacebuilding itself, deepening their understanding of how language contributes to harm.

The study also challenges the view of annotation as purely technical labour. Disagreements over labels required participants to debate intent, context, and historical meaning, mirroring core peacebuilding practices of dialogue and negotiated interpretation. By embedding local knowledge into the classifier, practitioners reclaimed interpretive authority over how their communities and conflicts were represented in data. In this sense, participatory annotation was not only a method for improving model performance but also an intervention that embedded peacebuilding norms within the analytical pipeline.

Nevertheless, improved technical performance does not necessarily translate into improved peace outcomes. A text classifier does not resolve conflict; it assists in navigating complex informational terrain. It can be understood as a customised compass that helps identify signals within large volumes of data. A more accurate compass does not determine the traveller's destination, nor does it guarantee success, but it can improve orientation. The core contribution of this study lies in demonstrating that when peacebuilders participate in calibrating that compass, deciding what it should detect and how its limits are understood, the resulting tool is better suited to the landscape in which it will be used.

# Recommendations

The findings from this study point to several practical recommendations for organizations considering participatory AI development in peacebuilding and comparable high-stakes contexts.

**Involve practitioners early.** The most significant gains in both model quality and practitioner ownership came from involving peacebuilders at the problem definition stage, not merely as validators of a finished product. Organizations commissioning AI tools for conflict analysis should build co-design into the project structure from the outset, including decisions about what to detect, how categories are defined, and what the tool will and will not be used for.

**Treat annotation as a deliberative process.** Disagreements between annotators should not be resolved by majority vote or discarded as noise. This study found that the sessions convened to work through labeling conflicts were among the most valuable parts of the process, producing both stronger training data and deeper practitioner understanding. Organizations should budget time and facilitation for structured disagreement resolution as a core component of annotation work.

**Bridge the gap between annotation and model development.** Practitioners in both cohorts understood the front end of the process well but remained uncertain about how their annotations were used in training. This gap limited technical capacity building and was experienced as a missed opportunity. Future projects should create structured touchpoints between practitioners and data scientists at the model development stage, not only during annotation.

**Plan for the psychological demands of annotating conflict content.** Sustained exposure to hateful and polarizing content had real effects on annotators, particularly in the Sudan cohort, where participants were processing material from an active war in their own country. Organizations should build in wellbeing support, limit session lengths, and create space for participants to process what they are reading.

**Invest in localization as a continuous practice.** Language in conflict contexts shifts deliberately and rapidly to evade detection. A classifier trained today may be outdated within months as coded terminology evolves. Organizations deploying these tools should plan for ongoing annotation cycles and model updates, and maintain relationships with practitioner communities who can identify emerging linguistic patterns.

**Make tools open and accessible.** Both models developed through this study are publicly available on HuggingFace. Peacebuilding organizations operating in low-resource contexts should not have to rebuild what already exists. The field benefits when tools are shared openly, and open publication also invites scrutiny that improves quality over time. The Kenya model was published in November 2025 on Hugging Face

(https://huggingface.co/datavaluepeople/Polarization-Kenya) and the Sudan classifier was published in January 2026 (https://huggingface.co/datavaluepeople/Hate-Speech-Sudan-v2).

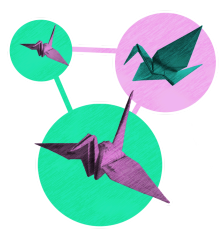